%% file: gomez_main.tex
\documentclass[multphys,vecphys]{svmult}

\usepackage{makeidx}     
\usepackage{graphicx}    
\usepackage{multicol}    
\usepackage{color}

\makeindex             

\begin{document}

\title*{Wave Modes in Collapsar Jets}
\author{Enrique A. G\'{o}mez \and
Philip E. Hardee}
\institute{Department of Physics and Astronomy\\ University of Alabama Box 870324, Tuscaloosa, AL 35487-0324. USA.\\
\texttt{enrique.gomez@ua.edu, hardee@athena.astr.ua.edu }}
%
%
\maketitle

\begin{abstract}

Collapsars may be a source for the ``long'' Gamma Ray Bursts (GRBs) in the BATSE catalog. Collapsars may radiate gamma rays anisotropically by beamed jet emission close to the observer's line of sight. These jets must penetrate the radiation-dominated medium of their collapsar progenitor and break through its atmosphere in order to produce a GRB. We present a study of previously published, axisymmetric, collapsar jet simulations. Here we use the linearized, relativistic fluid equations to find the Kelvin-Helmholtz modes that are triggered by recollimation shocks within the jet.  The modes will grow as they propagate with the jet. These are of interest since the light curves of GRBs show evidence of a variable flow from the GRB engine. We also evaluate effects of grid scaling in the numerical simulation and show that short wavelength modes are suppressed by grid scaling before the jet breaks out of the Helium shell.

\end{abstract}

\section{Introduction}
\label{sec:1}

The close association between GRB030329 and SN2003dh proves that a subset of ``long'' GRBs is associated with core collapse supernovae [1,2]. The light curve of GRB030329 steepens over time. Price et al. [3] interpret this as a broadening of a jet-like outf\mbox{}low from the GRB source. Multiwavelength observations by Uemura et al. [4] along with those from GRB021004 [5] strengthen the case for a variable jet outf\mbox{}low from the central engine.\\

These observations support the prompt (Type I) collapsar model of GRBs [6,7]. In this model, a progenitor star with mass $>$ 30 $\mathrm{M}_\odot$\space f\mbox{}ails to eject its envelope after the core collapses into a black hole. Accretion of free f\mbox{}alling material may drive a relativistic jet of gas that may break through the stellar envelope. At issue is whether hydrodynamic instabilities within the jet will grow so that the radial profile of jet velocity becomes time and space dependent. Such information could of\mbox{}fer insight into the production of gamma rays as well as prompt optical afterglows of GRBs.\\

\section{Dispersion Relations for Collapsar Jets}
\label{sec:2}

Analyses of collapsar jet models have shown that the jet is unstable as it propagates  through its progenitors atmosphere. Axisymmetric, relativistic hydrodynamic simulations by Aloy et al. (2000) [8] show evidence of pinch-body modes triggered by recollimation shocks. \\

We studied the simulations that Aloy et al. (2000) [8] made using the author's GENESIS code. In the radial direction, the energy deposition region extends from the inner grid boundary located at 200 km to a radius of 600 km (2-6 x 10$^{7}$ cm) . We studied two of their jets with constant energy deposition rates of dE/dt = 10$^{50}$ (C50) and 10$^{51}$ ergs s$^{-1}$ (C51). Our approach to analyzing the space-dependent structures is to use the linearized relativistic f\mbox{}luid equations to f\mbox{}ind the Kelvin-Helmholtz modes that may be triggered by recollimation shocks within the jet. Space-dependent perturbations propagate with a dispersion relation given in the appendix of Hardee, Clarke \& Rosen (1997) [9]. This dispersion relation assumes uniform conditions inside and outside a jet with sharp discontinuity at the jet surface.\\

The jets in the simulations have velocity profiles in the transverse direction. In order to identify modes in the simulation we must determine typical sound speeds in the jet and the external medium. To calculate the value of the relevant parameters in the jet and the external medium, we define the jet radius $R_j$ wherever $\Gamma v_{r}/c=0.5 \Gamma^{*} v_{r}^{*}/c$. Here $v_{r}$ is the radial velocity, $\Gamma = 1/\sqrt{1-v_{r}^2/c^2}$ is the Lorentz factor, $v_{r}^{*}$ is the maximum value for the jet velocity at a given cross section and $\Gamma^{*}$ the corresponding Lorentz value. Figure 1 shows the angular cross sections of pressure, density and velocity for the jet C50 near its base and near the head of the jet. We calculate the weighted average for jet speed and sound speed in the region $R\leq R_{j}$ and sound speed in the interval $R_{j} \leq R \leq 3R_{j}$. The local sound speed for a relativistic gas is def\mbox{}ined by
\begin{equation}
a_{s}\equiv\left\{ {\widehat{\gamma} p \over \rho +[\widehat{\gamma}/(\widehat{\gamma}-1)]p / c^2  }\right\}^{1\over 2}
\end{equation}
where $\widehat{\gamma}$ is the adiabatic index ($4/3$ for a relativistic gas), $p$ is the pressure, $\rho$ is the density, and $c$ is the speed of light. From the average of the jet speed typical values of $\Gamma$ range from 1-5; however, the C51 jet late in the evolution near the jet head reaches a $\Gamma \sim$ 20.

\begin{figure}
\centering

\includegraphics[height=8cm]{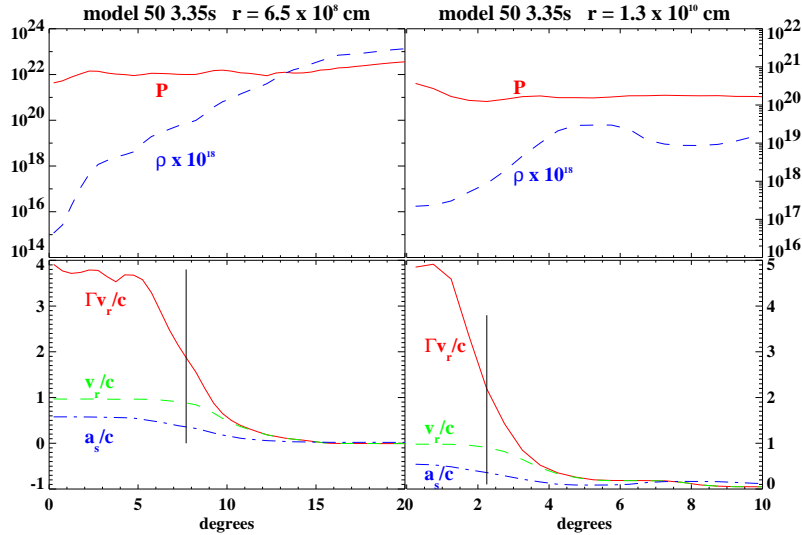}

%
%
\caption{Cross sections for pressure P, density $\rho$, velocity v$_{r}$ , and sound speed a$_{s}$ for the C50 jet at t=3.35s for cuts near the base and the jet head. Black vertical lines identify the defined jet radius, R$_{j}$.}
\label{fig:1}       
\end{figure}

\section{The Stages of Jet Evolution}
\label{sec:3}

We identify two distinct stages of jet evolution: downstream from the inlet and upstream from the breakout. The inlet region corresponds to where the jet is injected. In the C50 jet simulation at time t=3.35 s, we can study the stability downstream from the inlet near its base ($r\sim 10^{8} cm$). Breakout through the Helium shell of the progenitor occurs at $r\sim$ 10$^{11}$cm. Here the stability can be studied in the region immediately upstream from the jet head ($r\sim 10^{10} cm$). The dynamics of these two regions are comparable to the same jet at a later time and for the C51 simulation. Figure 2 shows the result of our stability analysis for these two regions with our calculated values for normalized wavelength and frequency for the fundamental, 1st, 2nd, and 3rd pinch body modes.\\

%
%
\begin{figure}
\centering

\vskip.2in
\includegraphics[height=4.5cm]{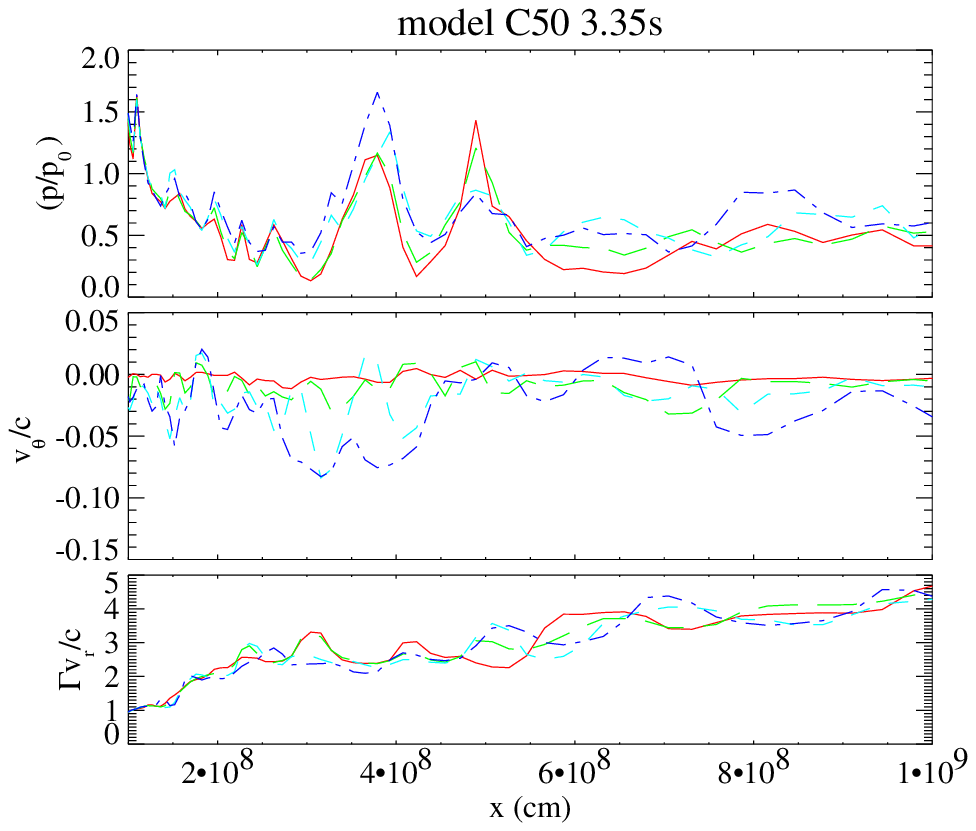}
\includegraphics[height=4.5cm]{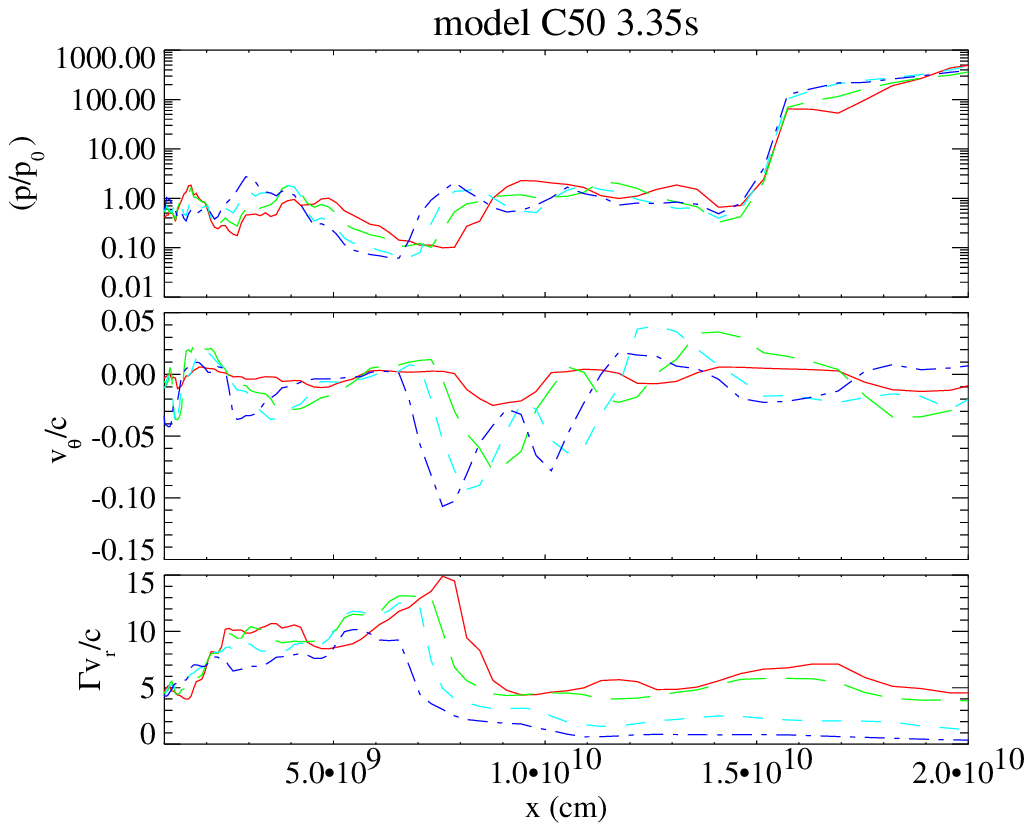}
\\
\vskip.2in
\includegraphics[height=4.5cm]{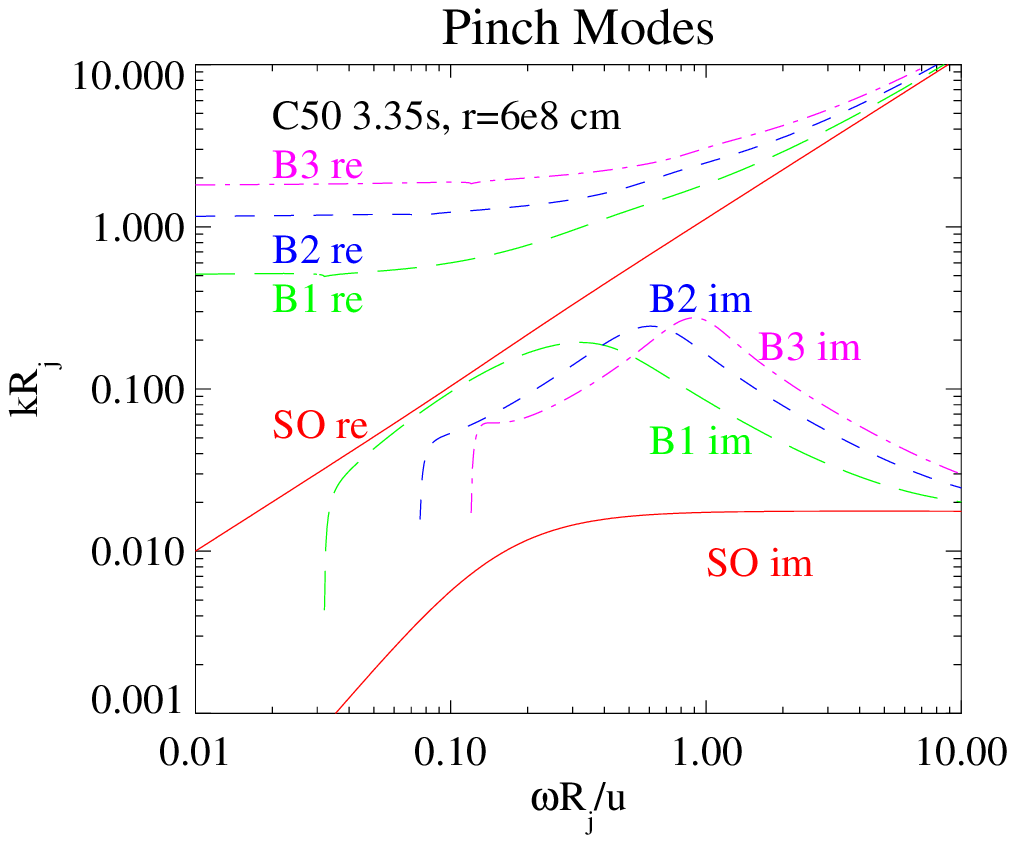}
\includegraphics[height=4.5cm]{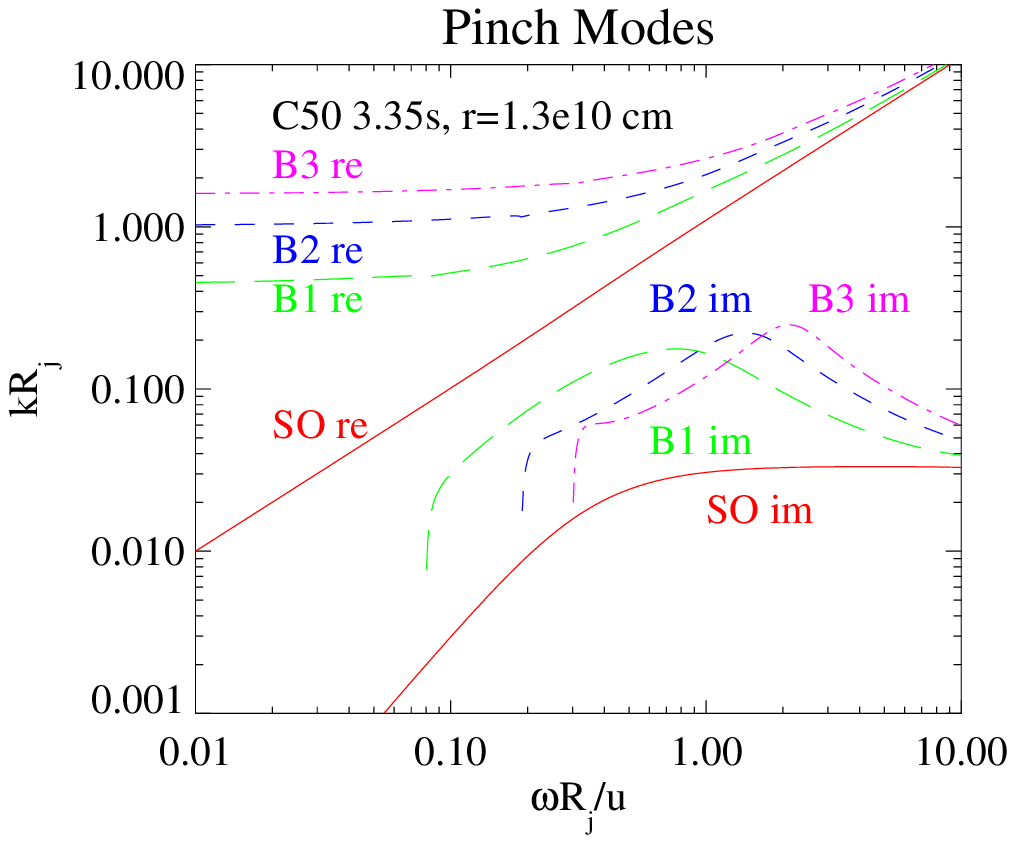}

%
%
\caption{Top row diagram shows cuts for the C50 jet at t=3.35 seconds for the jet close to the base (6.5 x 10$^{8}$ cm) and for the jet close to the head (1.3 x 10$^{10}$ cm). Cuts are for the angles 0.25$^{\circ}$, 1.125$^{\circ}$, 2.25$^{\circ}$ and 3.25$^{\circ}$ (\textcolor{red}{solid}, \textcolor{green}{long dash}, \textcolor{cyan}{short dash} and \textcolor{blue}{dash-dot} correspondingly). On the bottom row diagrams show the fundamental (\textcolor{red}{SO}), 1st, 2nd, and 3rd pinch body modes (\textcolor{green}{B1}, \textcolor{blue}{B2}, \textcolor{magenta}{B3}). These are the real and imaginary solutions for these regions.}
\label{fig:2}       
\end{figure}

Once the jet has penetrated the He shell, it enters a hydrogen medium where it will be overpressured. The jet will expand adiabaticaly in the perpendicular direction to the jet axis at $c/\sqrt{3}$ in the reference frame of the fluid [10,11].  Recollimation shocks below the He-H boundary shell destroy all upstream perturbations. At the same time they can trigger pinch modes downstream. We are most interested in perturbations just upstream of the jet head after it has crossed the He layer at 10$^{11}$cm. It is at this point that the jet becomes a causally disconnected wind where density and velocity perturbations ``freeze'' in the base of the wind.\\ 

\section{Conclusion}
\label{sec:4}

We solved the dispersion relations in Hardee, Clarke, \& Rosen 1997 [9] for $k(\omega)$ with the appropriate parameters from each cut of the jets. We found the solutions for the fundamental and the first three body modes for both jets. Figure 2 shows the solutions for C50 just downstream from the inlet and upstream from breakout at $t$=3.35s. We compute the wavelengths at the maximum growth rate $\lambda^{*}=2\pi v_{w}^{*}/\omega^{*}$ (where $v_{w}^{*}$ and $\omega^{*}$ are the wave speed and frequency at maximum growth) and the minimum growth length $\ell^{*}=(k_{I}^{*})^{-1}$ (where $k_{I}^{*}$ is the imaginary wave number). The growth time $\tau=\ell/v_{w}$ for any given solution is calculated from,

\begin{equation}
v_{w}=\frac{\omega k_{R}}{k_{R}^{2} + k_{I}^{2}}
\end{equation}

where $k_{R}$ is the real wave number. We find that our solutions depend on the distance from the inlet $r$. The range for maximum growth wavelength $\lambda^{*}$ for the body modes near the jet inlet is 0.1$r$ to $r$, and the range upstream from breakout is 0.06$r$ to 0.6$r$. The range for the growth length $\ell^{*}$ is 0.5$r$ to $r$ near the inlet, and for near breakout the range is 0.2$r$ to 2$r$. Near breakout the growth time $\tau$ ranges from 0.3 $r/c$ to 3 $r/c$ or 1 to 10 seconds at the distance of the He shell.\\

Our numerical analysis of jet stability assumes a sharp lateral boundary between the jet and the external medium. The analytical approach of Aloy et al. (2002) [12] for the same jet asumes an extended shear layer. For Lorentz factors of order 10, they predict a growth time $\tau\sim 0.01r/c$ which is one order of magnitude smaller than our method; however, the characteristic order of $\lambda^{*} \sim 0.1r - 0.5r$ is also what we predict.\\

One objective of our study was to find if there were wave modes that would not show up in the simulation because their length scale would be smaller than the computational grid scale. Aloy et al. (2000) use a logarithmic grid scaling so that the jet is better resolved at the base than at the head of the jet [8]. For the jet downstream from the inlet, grid scaling is $\sim$ 0.1 $R_{j}$ and this is suf\mbox{}ficient to resolve wave modes. Near the jet head the grid scaling is $\sim R_{j}$ and this can suppress the pinch body modes numerically. Since this region will be the most relevant to the evolution of the jet f\mbox{}low at breakout, simulations with greater grid resolution need to be done.\\

E. A. G. wishes to acknowledge the funding of this study through the Alabama Space Grant Consortium project. The authors would like to thank Miguel Angel Aloy for his generosity in sharing his simulation results and his comments.\\

%
\input{gomez_ref}


\printindex
\end{document}

%% file: gomez_ref.tex
%
%
%
%
%

%
%